\documentclass{aa}
\usepackage{psfig,latexsym,longtable}

\def\x{$\times$}
\def\>{$>$}
\def\<{$<$}

\def\newline{\hfil\break}
\def\mincir{\ \raise -2.truept\hbox{\rlap{\hbox{$\sim$}}\raise5.truept  
\hbox{$<$}\ }}                        % minore o circa uguale
\def\magcir{\ \raise -2.truept\hbox{\rlap{\hbox{$\sim$}}\raise5.truept  
\hbox{$>$}\ }}                        % maggiore o circa uguale

\begin{document}

\thesaurus{03(11.09.1 Cartwheel,11.19.3,13.25.2)}
\pagenumbering{arabic}
\title{
High energy phenomena in the Cartwheel's ring  
}

\author{Anna\,Wolter, Ginevra\,Trinchieri and Angela\,Iovino}

\institute{
Osservatorio Astronomico di Brera, via Brera 28, 20121
 Milano Italy}

\maketitle

\begin{abstract}

We present the first X-ray detection of the Cartwheel's ring.
The impact that created the characteristic optical structure is responsible
also for an enhanced level of star formation to which we can ascribe
the relatively high level of X-ray emission observed.
Deeper images and spectroscopic information are needed to confirm
or disprove the hypothesis that High Mass X-ray Binaries contribute
most of the radiation in this band.
\end{abstract}

\section{Introduction}
The origin of ring galaxies has been described by Lynds and Toomre (1976)
and Toomre (1978): 
a nearly head-on collision between a compact galaxy and a gaseous disk
creates an extra inward gravitational force that causes the orbits of
disk particles to contract.
The gravitational rebound that follows generates a transient density wave that
propagates outward, sweeping the interstellar gas and forming a rather
symmetrical feature, the ring, generally at 8-10 kpc (Theys and Spieghel 1977,
Hernquist and Weil 1993).
The presence of nearby companions, often with disturbed appearance, supports 
the collision theory for the formation of rings (Theys and Spieghel 1976, 
Hernquist and Weil 1993).
The gas swept up by the density wave and collected in the ring reaches
high densities and star formation phenomena are induced and enhanced. 
Evidence of high levels of star formation in rings is based on their blue 
optical colors, large far-infrared luminosities and on the presence
of large, luminous HII\ regions 
with a young population, consistent with an estimated age of rings of
$\sim 10^8$ yrs (Fosbury \& Hawarden 1977, Appleton \& Struck-Marcell 1987, 
Higdon 1995).
A comprehensive review of the properties of ring galaxies is given
by Appleton \& Struck-Marcell (1995).

One of the most spectacular and famous 
example of a recently formed ring galaxy is the A0035-324  system, 
named ``Cartwheel" because of its optical appearance: a 
bright, sharp outer ring linked by spiral spokes to the nucleus, 
reminiscent of a cartwheel.  
The Cartwheel is well studied in various wavelength ranges, but 
its X-ray properties have not been studied yet.
We present here X-ray HRI ROSAT observations of this peculiar object.

\section{The Cartwheel: a spectacular ring galaxy}

The Cartwheel is located in a tight, compact group (at z $\sim 0.03$)
of about 0.2 Mpc\footnote{We use H$\rm_0$ = 50 km s$^{-1}$ Mpc$^{-1}$, 
that implies a scale of 0.834 kpc/arcsec at the distance of the group, 
throughout the paper.}
physical size, found also by the newly developed
algorithm that defines the catalog of Southern Compact Groups 
(Iovino et al, in preparation). 
The group is composed of four members, with velocities all within 
400 km s$^{-1}$ from one another (Taylor and Atherton 1984; naming as in
Higdon 1995):
the Cartwheel, two galaxies $\sim 1^{\prime}$ NE (G1+G2) which are
probably in interaction (Higdon 1996), and a fourth object (G3)
at $\sim 3^{\prime}$ to the North.   The three companions 
are all of comparable magnitude (m$_B \sim 16$).
While it is likely that one of these is responsible for
the creation of the Cartwheel's ring, the identity of the intruder is 
still debated.
Higdon (1996) suggests  G3, from the trail in HI that seems to connect
it to the Cartwheel. However, recent simulations  by Athanassoula, Puerari
and Bosma  (1997) successfully reproduce the spokes structure 
assuming G2 as the intruder.

Two rings have been produced by the impact.
The outer one has the largest linear diameter measured in ring galaxies:
80$''$ along 
the major axis, corresponding to  $\sim$65 kpc at the distance of 180 Mpc.
The inner ring, close to the core, has been studied recently in detail with HST 
(Borne et al 1996; Struck et al. 1996): it is elliptical in shape with obvious
dust lanes crossing it. 

Many detailed observations of the Cartwheel are available, 
ranging from radio line (Higdon 1996) and continuum (Higdon 1996), 
to near- (Marcum et al., 1992) and far-infrared (Appleton and Struck-Marcell,
1987), optical (Theys and Spiegel 1976, Fosbury and Hawarden 1977) and 
H$\alpha$ images (Higdon 1995) and line spectroscopy (Fosbury and 
Hawarden 1977). All have confirmed
the presence of a recent starburst in the outer ring, without any
corresponding activity in the inner ring, nucleus or spokes, believed
to be relatively devoid of gas.   Most 
of the activity is confined in fact in the S-SW portion of the ring,
where massive and luminous HII regions characterized by large
H$\alpha$ luminosities and equivalent widths are found (Higdon 1995).
Both dynamical considerations and stellar evolution models suggest 
an age of 2-4 \x 10$^8$ yr for the star burst. The estimated supernova
rate, as high as 1 SN/yr (i.e. almost two orders of magnitude higher than in 
normal galaxies), coupled with the 
evidence of very low metallicity, measured
in O, N and Ne, 
also supports the view that star formation in the ring is a
recent phenomenon and that gas currently forming stars 
was nearly primordial at the time
of the impact (Fosbury \& Hawarden 1977; Higdon 1995; Marcum et al 1992).  

\section{The X-ray picture} 

We have detected and imaged the Cartwheel for the first time in the X-ray band,
using the HRI on board ROSAT (Tr\"umper 1983). 
The observation (ROR= 600748) was carried out 
in December 1994, for a total ontime of 60225 sec.
Prior to our observation, only a very uncertain upper limit, derived
from a short $Einstein$ IPC exposure, was available (Ghigo et al. 1983).

The data have been analyzed with the {\it xray} package in \verb+IRAF+.
We first smoothed the image with different Gaussian widths to enhance
the significance of the features in the image and inspected the results
to check the accuracy of the absolute positions in the sky,
namely the location of the X-ray emission with respect to the 
target galaxy.

\begin{figure}[h]
\psfig{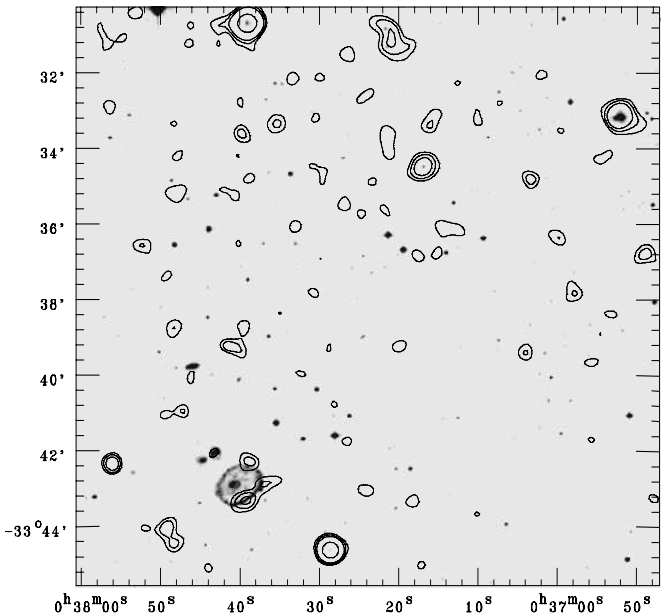}
\caption{The large field shown here indicates the coincidence of
several optical counterparts with X-ray sources in the field of the
Cartwheel. The HRI data are smoothed with a Gaussian with 
$\sigma=10^{\prime\prime}$.
}
\end{figure}

Figure 1 shows the superposition of the X-ray data smoothed with a 
Gaussian function with  $\sigma=10''$ on the Digitized Sky 
Survey plate\footnote{The Digitized Sky Survey was produced at the Space Telescope 
Science Institute (STScI) under U.S. Government grant NAG W-2166.}.
The positioning of the X-ray emission is accurate, since several X-ray 
point sources coincide with their optical counterparts in the field of 
the Cartwheel. Searching the NED database we find that the source at 
RA(J2000) 00$^{h}$ 36$^m$ 52$^s$
and Dec(J2000) $-33^{\circ}$ 33$^{\prime}$ 12$^{\prime\prime}$ 
coincide with ESO 350-IG 038 (a group or interacting pair containing a Seyfert
galaxy) and the source at RA(J2000) 00$^{h}$ 37$^m$ 29$^s$
and Dec(J2000) $-33^{\circ}$ 44$^{\prime}$ 40$^{\prime\prime}$ with
the radio source PKS 0035-340.

\begin{figure}
\psfig{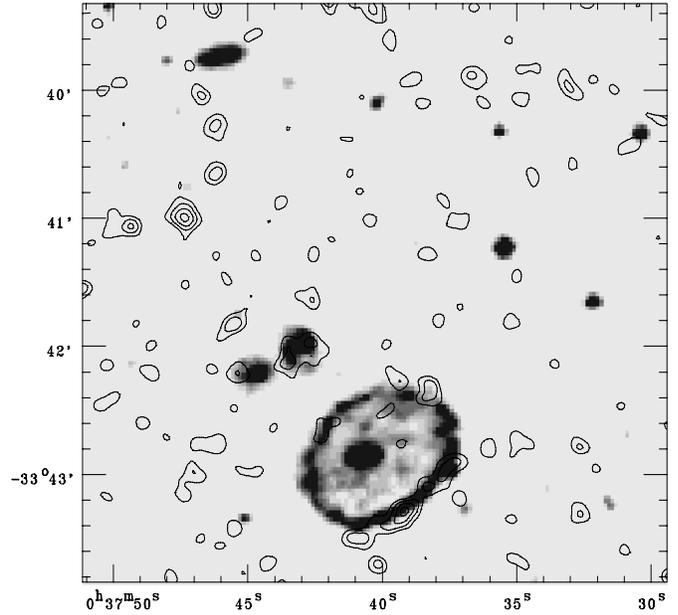}
\caption{X-ray emission from the Cartwheel galaxy and the 3 companions
superposed on the Digitized Sky Survey plate.  Isocontours are at
0.041, 0.076, 0.116 and 0.156 cts/arcsec$^2$  over the background,
assumed at 0.054 cts/arcsec$^2$ (see text).
The HRI data are smoothed 
with a Gaussian function with $\sigma = 3''$. }
\end{figure}

In order to improve the S/N ratio we have made a selection in Pulse Height
Analyzer (PHA) channels.
We have extracted total counts from a 
2$^{\prime}$ diameter circle centered on the Cartwheel optical position,
and plotted their distribution in PHA channels
against the distribution of background counts, extracted in a nearby 
region devoid of sources. 
Selecting PHA channels 2-10
reduces the background considerably (25-30\%),
while retaining a large enough range in PHA channels 
that changes in the 
gain across the detector should not significantly affect
our results. This PHA selection applies to all further
results.

Figure 2 shows the X-ray emission from the Cartwheel galaxy and its 
companions, again superposed on the DSS.   The data are 
smoothed with a Gaussian function with $\sigma = 3''$, that matches the 
HRI spatial resolution, so that the smallest resolvable features are
of the order of 3 kpc at the distance of the group.
As can be seen, all of the emission is located in the outer ring, while the 
nucleus, inner ring and spokes are not detected.
In particular the emission is stronger in the Southern quadrant, where 
the H$\alpha$ emission is also stronger 
(Higdon 1995).

\begin{figure}
\psfig{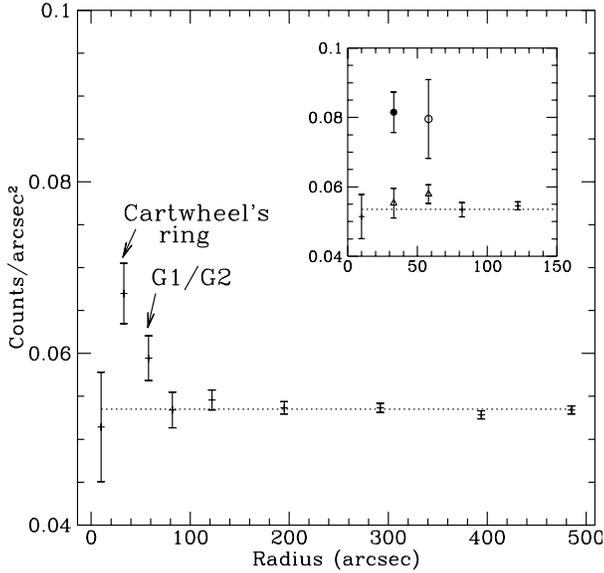}
\caption{Radial distribution of the TOTAL counts detected in the
Cartwheel HRI observation, centered on the optical nucleus. The annuli
that contain most of the optical ring and the G1/G2 pair are indicated. 
The dotted line indicates the background derived from the average of points
at radii $\geq 150^{\prime\prime}$.
INSET: enlargement of the internal region of the profile. The
solid and open circles indicate respectively the azimuthal regions
containing the Cartwheel's ring and the G1/G2 pair, while
triangles indicate the complementary regions in the annuli. 
See text for details.
}
\end{figure}

To best define the location and extent of the X-ray emission 
relative to the optical one and to determine
the background level we have derived the radial distribution of the total 
counts in the image by summing over all angles in radial bins
centered at RA(J2000) 00$^{h}$ 37$^m$ 41.1$^s$
and Dec(J2000) $-33^{\circ}$ 42$^{\prime}$ 54.0$^{\prime\prime}$ (the optical
nucleus of the Cartwheel),
after removal of unrelated point sources.
We have chosen the inner bins such that they maximize the S/N of the
emission, while the outer bins are at regular intervals of 1.6$^{\prime}$
and are representative of the background level.
The resulting profile is plotted in Figure 3. At distances greater than
150$^{\prime\prime}$ the profile is flat
with an average value of 0.054$\pm 0.001$ cts/arcsec$^2$.

Figure 3 confirms the conclusion of the visual inspection 
of the image, that all of the emission is in an annulus at 
$\sim 0.4^\prime$-1.1$^\prime$ from the nucleus, and can be attributed to
the outer ring of the Cartwheel (the majority) and to the two nearmost
companions  G1/G2. 
No excess emission is detected inside the ring, or 
associated with the group or galaxy G3.
To better show the significance of the X-ray emission related to the 
Cartwheel's ring and the pair G1/G2 we plot in the inset of Figure 3
an enlargement of the profile in different azimuthal sectors.
We have chosen angles $170^{\circ} <\theta <330^{\circ}$ and 
$20^{\circ} <\theta <45^{\circ}$
(from North to East) and their complementary regions for the 
24$^{\prime\prime}$ - 46$^{\prime\prime}$
and 46$^{\prime\prime}$ - 70$^{\prime\prime}$ annuli respectively,
while all other points are azimuthally averaged over all angles. 
The comparison between the regions of emission (solid and open dots in the
inset) and their complementary regions at the same angular distance
(triangles) clearly indicates the association between X-ray and optical 
emission
in both the Southern part of the ring and the G1/G2 galaxies.
The complementary regions are consistent with the 
background level.

To compute the flux from the Cartwheel's ring
we extracted the net counts in an elliptical annulus matching the 
X-ray emission from the outer ring
for a total of 115.7$\pm$22 net counts above the derived
background.
We also detect 34.3$\pm$11.5 counts from en ellipse enclosing
the G1/G2 galaxies.
Given the location of the sources, near the center of the field of view,
vignetting corrections are very small and are therefore ignored.
The scantiness of the counts and the extremely limited spectral capabilities 
of the HRI prevent us from obtaining proper spectral information.
We have assumed a number of different models, from
a $raymond$ spectrum (hot gas) with kT= 0.5 -- 5 keV and cosmic abundance
20\% or 100\%, to a $power~law$ (stellar sources) with $\alpha$ = 0.5 -- 2.
and low energy absorption corresponding to the line-of-sight
column density N$_{H}= 2.0 \times 10^{20}$ cm$^{-2}$ (from Dickey \& Lockman,
1990).
The results of the various models are virtually insensitive to the choice 
of parameters within the HRI band.  We therefore
use a mean conversion factor for which the detected count rate corresponds
to a flux f$_x = 6.5 \times 10^{-14}$ erg cm$^{-2}$ s$^{-1}$ and
to a luminosity L$_x = 2.5 \times 10^{41}$ erg s$^{-1}$ for the
Cartwheel's ring and to a flux f$_x = 1.9 \times 10^{-14}$ erg cm$^{-2}$ 
s$^{-1}$ and a luminosity L$_x = 7.4  \times 10^{40}$ erg s$^{-1}$ 
for the galaxy pair G1/G2. 
The luminosity of the ring could be as high as L$_x  \sim  5.0
\times 10^{41}$ erg s$^{-1}$ if an intrinsic absorption 
of N$_{H}= 2 \times 10^{21}$ cm$^{-2}$ (corresponding to the reddening
observed in the HII regions) is assumed. 

A small fraction (of about 1/10) of the flux attributed to the 
Cartwheel might be associated with a
star superposed on the outer ring (well visible in the HST image
[NASA Press Release PRC96-36a, Nov. 26, 1996] 
and already identified as an object with the colors of
a G star of 17-18 mag, Marcum et al. 92).  
The X-ray to optical flux ratio, estimated in a small circle at the
star position,
is consistent with values of G stars of this magnitude
(see nomogram in Maccacaro et al. 1988).

The upper limit for point source detection at the distance of the
Cartwheel is of the order of $10^{40}$ erg s$^{-1}$.
This is higher than what expected from
a starburst nucleus (usually a few  $\times 10^{39}$ erg s$^{-1}$). Therefore
we cannot exclude that the nucleus presents a modest activity, below
the detection limit.
Also the galaxy G3, or the area corresponding to the HI bridge between 
the Cartwheel and G3 might show
some X-ray emission at a level too faint to be detected by this observation.
Since the galaxies belong to a (compact) group we
have searched for extended emission from the intragroup hot
gas: integrating over different areas does not result in a significant 
detection at the 3$\sigma$ level.
We derive a  3$\sigma$ upper limit
of  L$_X \leq 6 \times 10^{42}$ erg s$^{-1}$ on the level of the
emission from the group, within a radius of
4$^\prime$, not stringent enough to exclude 
the presence of hot gas in the area.

\section{Results }

The ROSAT HRI observation shows that the X-ray emission is linked
to the Cartwheel's ring, while the nucleus and the interior of the
ring are not detected. 
The G1/G2 pair is detected at a relatively high X-ray luminosity.
Higdon (1996) gives B magnitudes for the 2 objects:
the G2 is an elliptical galaxy of L$_B = 4.8 \times 10^9 L_{\odot}$ for which 
we expect an X-ray  luminosity in the range $1.5 \times 10^{39}$ --
$5 \times 10^{40}$
erg s$^{-1}$  and G1 is a spiral of L$_B = 6. \times 10^9 L_{\odot}$
for which we expect a few $\times 10^{39}$ erg s$^{-1}$ 
(following Fig. 19 in Fabbiano et al. 1992 for the L$_X$ vs. L$_B$
relationships).
The observed luminosity therefore appears higher than expected on the
basis of observed properties of normal galaxies. However,
the pair is believed to be interacting as suggested by HI observations 
(Higdon 1996) and thus a higher emission is not unlikely.
Furthermore, since the number of detected counts is too small to establish
if the X-ray emission is extended or not, we cannot exclude the presence
of a small active nucleus in one of the two galaxies.

No emission is detected from the intragroup gas, albeit with 
a non-stringent upper limit. The HRI however is not very sensitive
to low surface brightness emission.

A relatively high X-ray luminosity is detected from the Cartwheel,
about a factor 10-30 higher than expected for a normal galaxy of the
same optical luminosity (from the relationship derived in Fabbiano et
al. 1992).   The HRI image reveals a remarkable coincidence between the
X-ray and the $H\alpha$ emission in the ring (see Higdon 1995, figures
2a-2c-2e), indicative of a close association with the star formation
phenomenon.  X-ray emission is higher in the area that contains the
brightest HII regions (S-SW quadrant), while no emission is observed
from the nucleus or the inner ring, where no star formation enhancement
is detected from optical/radio/IR data.  This HRI observation does not
suggest a one-to-one correspondence between the location of the X-ray
peaks with the bright H$\alpha$ knots or the radio continuum sources.
However, given the limited statistics, we have to postpone a reliable
comparison of the details of the morphology at different wavelengths to
future more sensitive observations.

While the general association with the ring is firmly established, 
the origin of the X-ray emission remains to be fully understood. 
We explore
different possibilities, based on the knowledge of emission in
similar environments. 
No ring galaxy has been detected yet in the X-ray band with
a data quality that allows a firm determination of the emission 
mechanisms. 

From the relationship between the observed IR luminosity
and the X-ray emission for starburst galaxies (eg. David et al. 1992, Boller
et al. 1992) we expect L$_{X}$ due to the total starburst
contribution in the range $3-7 \times 10^{40}$ erg s$^{-1}$
for the observed IRAS luminosity of the Cartwheel 
L$_{FIR} = 1.4 \times 10^{10} L_\odot$ 
(Appleton \& Struck-Marcell 1987).
The X-ray luminosity observed is higher than this by a factor of a few, but 
smaller than what would be expected if the FIR flux were due to a Seyfert
nucleus (L$_{X} \sim 1 \times 10^{42}$ erg s$^{-1}$,
using the Boller et al. 1992 regressions for Sy nuclei). 
Moreover, the non-detection of a central (nuclear) source makes the 
discrepancy even larger. This
is an independent confirmation that the IR luminosity in the IRAS band
is associated with starburst and not to nuclear activity.

From the observed L$_{X}$ we can estimate some physical
quantities, assuming that the emission is all due to a
hot plasma.  The density
of electrons, for a temperature T=$1 \times 10^7 K$ is:
$$
N_e = \sqrt { L_X \over {2.4 \times 10^{-27} T^{1/2} V} } cm^{-3} = 1.2 \times 10^{-2}  cm^{-3}
$$
where the approximate volume is derived assuming that the detected emission 
in the South-West region comes 
from a cylinder of radius 7$^{\prime\prime}$ and height 72.5$^{\prime\prime}$ 
and is V = $ 1.76 \times 10^{68}$ cm$^3$.
The corresponding cooling time: 
$$
t_{ff} = {1.8 \times 10^{11} T^{1/2} \over N_e} sec = 4.7 \times 10^{16} sec = 1.5 \times 10^9 yr
$$
is longer than the burst lifetime.
The total mass of hot gas in the volume sampled is M$_{gas} \mincir
1.77 \times 10^9 M_\odot$, to be compared with  $9.3 \times 10^9 M_\odot$ 
in HI derived in the entire Cartwheel galaxy from 
radio measurements (Higdon 1996).

On the other hand, the Cartwheel contains a relatively high number of
young stars, supernovae and bright HII regions, resulting from the
recent burst of star formation that can contribute substantially to
the X-ray activity.
Estimates based on optical and infrared data suggest the presence of a
few $ \times 10^{6}$ O stars in the whole ring (Marcum et al. 1992,
Fosbury \& Hawarden 1977, Appleton \& Struck-Marcell 1987), implying
\mincir $10^7$ O-B type stars.  Even in the generous assumption that
all emit at the highest level of $\sim$ 10$^{33}$ erg s$^{-1}$ as
observed in our Galaxy (Vaiana et al. 1981; Chlebowski et al. 1989),
the latter would account at most for \mincir 5\% of the observed
luminosity.  However, High Mass X-ray Binaries (HMXB) are formed about
1 every 500 O star (Fabbiano et al., 1982); as a consequence, more
than 2000 HMXB could have formed in the Cartwheel.  HMXB have a range
of X-ray luminosities: in our Galaxy, the average L$_X$ is $\sim
10^{37}$ erg s$^{-1}$, while in the Magellanic Clouds (MC), is $\geq
10^{38}$ erg s$^{-1}$; the higher luminosity in the MC is believed to
be linked to the lower abundances found in the Clouds (van Paradijs
and McClintock, 1995).  If the HMXB in the Cartwheel are of the same
type found in the MC, they could contribute $\geq L_X \sim 2\times
10^{41}$ erg s$^{-1}$, consistent with what we observe.

A confirmation of the hypothesis that HMXB are the main source of the
X-ray emission from the Cartwheel will however only come from a
reliable measure of the X-ray spectrum spectrum coupled with a high
spatial resolution, that will become available with future X-ray
missions.

It is likely that the HMXB would be located in giant HII regions,
which would appear extremely bright compared to equivalent structures
in our Galaxy.  In fact, giant HII regions and complex structures,
typically coincident with peaks of H$\alpha$ emission, are observed in
actively forming objects like the interacting system ``The Antennae"
(Read et al. 1995; Fabbiano et al. 1997) with intrinsic X-ray
luminosities reaching several \x 10$^{40}$ erg s$^{-1}$.  
Around young binary systems superbubbles could also be formed, which
could be detected in X-rays as bright sources.  However, similar
systems observed in the LMC only reach luminosities of a few \x
10$^{37}$ erg s$^{-1}$ each (e.g. Wang \& Helfand, 1991).

Other phenomena might contribute to the X-ray emission of the
Cartwheel, however  we do not expect them to dominate on energetic
grounds.  We list them briefly here:

a) a phenomenon that typically heats the gas to X-ray emitting
temperatures are shocks; however, strong shocks are excluded from
radio and optical observations (Higdon 1996), while a shock
propagating at a speed equal to the expansion velocity of the ring
($\sim$ 60 km/s, from HI data; Higdon 1996, or even lower from recent
H$\alpha$ measurements; Amram et al. 1998) would heat the gas at a
temperature of kT $\leq 0.05$ keV, quite low to be detected by the
HRI. The detected counts would correspond in fact to the very
improbable luminosity of a few $10^{42}$ erg s$^{-1}$, for such a low
temperature.  The PHA distribution of the counts also argues against
such an extreme case: in spite of the very poor spectral capabilities
of the HRI, we would expect all the counts in the softest channel, for
such low temperature.  We however cannot exclude that shocks in high
density environments like giant HII regions give a significant
contribution.  As discussed above, only the much needed spectral
information will help us in this matter.

b) The ISM of the Cartwheel's precursor (thought to be a late spiral)
is unlikely, since it would have an X-ray luminosity at least
100\x higher than that
of the ISM in normal spiral galaxies (in the 10$^{39}$ erg s$^{-1}$
range at most; Fabbiano 1996).

c) Supernovae and Supernova remnants (SNR) are likely to be
responsible for at least a fraction of the observed emission.  Given
the mean X-ray luminosity of a SNR ($L_X(SNR) \sim 10^{36}$ erg
s$^{-1}$ over a typical time $\tau(SNR)\sim2\times 10^4$ yr; Cowie et
al. 1981) the total contribution from SNR during the burst is $\nu
\times \tau(SNR) \times L_X(SNR) = 4 \times 10^{40}$ erg s$^{-1}$ 
for a SN rate $\nu$ = 1 yr$^{-1}$ as measured by Fosbury \& Hawarden 
(1977).  This is a small fraction of the observed luminosity.

d) In spite of the relatively large number of young stars, coronal
emission from normal stars is unlikely to contribute significantly
to the observed L$_X$, as seen above.

Even if a spectral measurement with high S/N is necessary to
discriminate between the different contributions,
we can however check the rough estimates of the masses required by the
enhanced star formation scenario against other estimates. 

To derive the total mass in stars,
we can apply the results of David et al. (1992) for starburst galaxies,
namely that the efficiency of producing X-rays through star formation 
is roughly independent of the star
formation rate.  If about $10^{47.5}$ ergs of radiant energy are produced 
for each solar mass consumed into stars, we expect a rate
$ \dot{M} \sim 20 -  50 \ M_{\odot}/yr $. 
The range quoted reflects the
uncertainty on the intrinsic luminosity due to
absorption.
The value of $\dot M$ is consistent with previous 
estimates from e.g. H$\alpha$ 
measurements (Higdon 1995):     
11 $M_{\odot}/yr$ (high mass stars) -
67 $M_{\odot}/yr$ (total stellar population).

The total mass of stars formed, M$_{star} = \dot{M} \times t_{burst} =
0.5 - 2 \times 10^{10} M_\odot$ (taking into account the uncertainties
in both L$_x$ and $t_{burst}$), is also consistent with the estimates
of Fosbury \& Hawarden (1977), who find $\sim 2 \times 10^{10}
M_\odot$ assuming a mass-to-light ratio of the order of solar for the
extreme population I.  Using the H band luminosity from the data in
Marcum et al. (1992) and the total mass in stars derived above,
L$_H$/M $\sim 2.5-10 $, consistent with the value found for HII
galaxies by Oliva et al. (1995), thus strengthening the idea of a
strong link between X-ray emission and star formation activity in this
object .

\section{Summary}

The HRI observation of the Cartwheel points to a link of the X-ray
emission and the star formation activity also in an environment
created by a face-on encounter between two galaxies.  The observed
$L_X$ is higher than that expected from normal galaxies, and could be
explained by the (high) number of HMXB created by the starburst if
their X-ray luminosities are of the same order of the ones in the
Magellanic Clouds.  However a substantial contribution might come from
a number of other sources, including the gas heated by the star
formation related phenomena.  The overall parameters derived, like the
total mass of stars formed and the mass-to-light ratio, are in
agreement with previous results obtained with observations at other
wavelengths. However, a precise interpretation of the mechanism of
X-ray emission in the Cartwheel's ring is hampered by the lack of
spectral resolution and sufficient statistics.  Good quality spectral
data and more sensitive observations are therefore necessary in order
to reduce the uncertainties and to confirm our hypotheses.  The
upcoming AXAF and XMM satellites will provide us with the necessary
tools to disentangle the problem of the origin and mechanisms of the
high energy emission in this peculiar object and to add to our
understanding of the star burst phenomenon.

\begin{acknowledgements}
We thank Laura Maraschi, Catarina Lobo and Roberto Della Ceca
for useful comments.
GT thanks prof. Tr\"umper for hospitality at MPE while part of
this work was done.
This research has made use of the NASA/IPAC Extragalactic
Database (NED) which is operated by the Jet Propulsion Laboratory,
CALTECH, under contract with the National Aeronautics and Space
Administration (NASA). 
This work has received partial financial support from the Italian 
Space Agency.

\end{acknowledgements}

%
% ---- Bibliography ----
%

\end{document}